\documentclass{blois}

\def\be{\begin{equation}}
\def\ee{\end{equation}}
\def\bea{\begin{eqnarray}}
\def\eea{\end{eqnarray}}

\newcommand{\Photo}{}

\begin{document}
\vspace*{4cm}
\title{High-resolution SZ imaging of clusters of galaxies with the NIKA2 
camera at the IRAM 30-m telescope}

\author{F.~Mayet$^1$,
R.~Adam$^{1,2}$,
P.~Ade$^3$,
P.~Andr\'e$^4$,
M.~Arnaud$^4$,
H.~Aussel$^4$,
I.~Bartalucci$^4$,
A.~Beelen$^5$,
A.~Beno\^it$^6$,
A.~Bideaud$^3$,
O.~Bourrion$^1$,
M.~Calvo$^6$,
A.~Catalano$^1$,
B.~Comis$^1$,
M. De Petris$^7$,
F.-X.~D\'esert$^{8}$,
S.~Doyle$^3$,
E.~F.~C.~Driessen$^9$,
J.~Goupy$^6$,
C.~Kramer$^{10}$,
G.~Lagache$^{11}$,
S.~Leclercq$^9$,
J.~F.~Lestrade$^{12}$,
J.~F.~Mac\'ias-P\'erez$^1$,
P.~Mauskopf$^{3,13}$,
A.~Monfardini$^6$,
E.~Pascale$^3$,
L.~Perotto$^1$,
E.~Pointecouteau$^{14}$,
G.~Pisano$^3$,
N.~Ponthieu$^{8}$,
G.~W.~Pratt$^4$,
V.~Rev\'eret$^4$,
A.~Ritacco$^{10}$,
C.~Romero$^9$,
H.~Roussel$^{15}$,
F.~Ruppin$^1$,
K.~Schuster$^9$,
A.~Sievers$^{10}$,
S.~Triqueneaux$^6$,
C.~Tucker$^3$,
R.~Zylka$^9$}

\address{$^{1}$ Laboratoire de Physique Subatomique et de Cosmologie, Universit\'e Grenoble Alpes, CNRS/IN2P3, 53, avenue des Martyrs, Grenoble, France\\
 $^{2}$Laboratoire Lagrange, Universit\'e C\^ote d'Azur, Observatoire de la C\^ote d'Azur, CNRS, Blvd de l'Observatoire, CS 34229, 06304 Nice cedex 4, France\\
 $^{3}$Astronomy Instrumentation Group, University of Cardiff, UK \\
 $^{4}$Laboratoire AIM, CEA/IRFU, CNRS/INSU, Universit\'e Paris Diderot, CEA-Saclay, 91191 Gif-Sur-Yvette, France\\
 $^{5}$Institut d'Astrophysique Spatiale (IAS), CNRS and Universit\'e Paris Sud, Orsay, France\\
 $^{6}$Institut N\'eel, CNRS and Universit\'e Grenoble Alpes, France\\
 $^{7}$Dipartimento di Fisica, Sapienza Universit\`a di Roma, Piazzale Aldo Moro 5, I-00185 Roma, Italy\\
 $^{8}$Institut de Plan\'etologie et d'Astrophysique de Grenoble, Univ. Grenoble Alpes, CNRS, IPAG, F-38000 Grenoble, France\\ 
 $^{9}$Institut de RadioAstronomie Millim\'etrique (IRAM), Grenoble, France\\
 $^{10}$Institut de RadioAstronomie Millim\'etrique (IRAM), Granada, Spain\\
 $^{11}$Aix Marseille Universit\'e, CNRS, LAM (Laboratoire d'Astrophysique de Marseille) UMR 7326, 13388, Marseille, France\\ 
 $^{12}$LERMA, CNRS, Observatoire de Paris, 61 avenue de l'observatoire, Paris, France\\
 $^{13}$School of Earth and Space Exploration and Department of Physics, Arizona State University, Tempe, AZ 85287\\
 $^{14}$Universit\'e de Toulouse, UPS-OMP, Institut de Recherche en Astrophysique et Plan\'etologie (IRAP), Toulouse, France\\
 $^{15}$Institut d'Astrophysique de Paris, Sorbonne Universit\'es, UPMC Univ. Paris 06, CNRS UMR 7095, 75014 Paris, France
 }

\maketitle\abstracts{
The development of precision cosmology with clusters of galaxies requires
high-angular resolution Sunyaev-Zel'dovich (SZ) observations. As for now, arcmin resolution SZ observations ({\it e.g.} 
SPT, ACT and Planck) only allowed detailed studies of the intra cluster medium for low redshift clusters ($z<0.2$).
With both a wide  field of view (6.5 arcmin) and a high angular resolution (17.7 and 11.2 arcsec at 150 and 260 GHz), 
the NIKA2 camera installed at the IRAM 30-m telescope (Pico Veleta, Spain), will bring valuable information in the field of SZ imaging of clusters of 
galaxies. The
NIKA2 SZ observation program will allow us to observe a large sample of clusters (50) at redshifts
between 0.4 and 0.9. As a pilot study for NIKA2, several clusters of galaxies have been observed with the pathfinder, NIKA, at the
IRAM 30-m telescope to cover the various configurations and observation
conditions expected for NIKA2.}

\section{The NIKA2 camera at the IRAM 30-m telescope}
NIKA2 is a millimetre camera~\cite{Calvo16,NIKA2-Adam}, made of Kinetic Inductance Detectors (KID) and operated at 150 mK, which 
has been installed in September 2015 at the focus of the IRAM 30-m telescope.
NIKA2 observes the sky at 150 and 260 GHz with a wide field of view (FOV), 6.5 arcmin (2896 detectors),  a high-angular resolution 
(17.7 and 11.2 arcsec, respectively),  and a state-of-the-art sensitivity (6 and 20 $\rm mJy.s^{1/2}$, respectively). 
The NIKA camera~\cite{Monfardini11,Bourrion12,Calvo13,Monfardini13,Catalano:2014nml}  was a pathfinder of NIKA2 that has been operated 
at the IRAM 30-m telescope from 2012 to 2015 with a smaller field of view  (1.8 arcmin) due to the reduced number of KIDs (356). 
The performance of these cameras at the IRAM 30-m telescope is described in~\cite{NIKA2-Adam,Catalano:2014nml}.

The NIKA2 camera is  well suited for high-resolution SZ observations of cluster of galaxies for several reasons.
First, it is a dual-band camera operating at frequencies for which the SZ signal is negative and slightly positive respectively. The
260-GHz map may be used for the detection for point sources or as a template of the atmospheric noise. Then, a high angular resolution and a large field of view~\cite{NIKA2-Adam} enable NIKA2 to well match intermediate and high redshift clusters. Moreover, as the NIKA2 field of view is about the size of the beam of the Planck's intruments, 
the combination of Planck and NIKA2 data will enable a SZ mapping of clusters of galaxies at all scales, 
from the core to the outskirts (up to $R_{500}$).
Eventually, the  sensitivity in Compton parameter units is expected to be of the order of 
$10^{-4}$ per hour and per beam, allowing us to obtain reliable SZ mapping at high signal-to-noise ratio in a few hours per cluster.

\section{The need for high-resolution SZ imaging of clusters}
The SZ effect~\cite{Sunyaev:1972,Sunyaev:1980vz,Birkinshaw:1998qp} is an inverse Compton scattering of CMB photons   
on hot electrons of the intra-cluster medium. It induces a shift of the CMB black-body spectrum to higher frequency, with 
a decrease of the  CMB intensity below $217 \ {\rm GHz}$ and an increase above. 
The SZ effect is thus a spectral distortion of the CMB spectrum. This is the reason why it is redshift-independent and can 
thus be used for the observation of high-redshift clusters. The SZ signal is the Compton parameter $y$, which is proportional to the electronic pressure $P_{\mathrm{e}}$ integrated along the line of sight, as  
$y \propto \int P_{\mathrm{e}} dl$. SZ observations provide imaging of the line-of-sight integrated pressure of the intra-cluster medium
(ICM). They can be combined with X-ray  observation (which surface brightness is related to the electronic density,  
 $S_X   \propto \int n_e^2 \Lambda(T_e, Z) dl$) in order to probe the clusters physical properties. 

Clusters of galaxies constitute powerful tools to study cosmology as their number and 
distribution in mass and redshift  is dependent of the geometry of the Universe.
Cosmological parameters~\cite{Ade:2013lmv,Ade:2015fva}  have been constrained 
by using cluster counts, as a function of 
redshift and mass, for a sample of clusters identified by their SZ  effect   
by  the Planck Satellite~\cite{Ade:2015gva,Ade:2013skr}, the Atacama Cosmology Telescope 
(ACT)~\cite{Hasselfield:2013wf} and the South Pole Telescope (SPT)~\cite{Bleem:2014iim}. However, it requires an estimation of the cluster halo mass, 
which is commonly obtained from X-ray observation assuming clusters are in hydrostatic equilibrium. In order to leverage large samples with 
inhomogeneous follow-up data and to estimate the selection function, a mass-to-observable scaling law is also needed to relate  
cluster observables, {\it e.g.} the integrated Compton parameter, with cluster parameters, such as the total mass. The calibration precision of this law is subject to our current understanding of the intra-cluster gas physics. Biases in the absolute mass calibration, such as a departure from hydrostatic equilibrium, which may increase with redshift, 
must be taken into account. The scatter in the mass-observable relation, linked to the intrinsic cluster physics, must also be included in the analysis.

Currently, there is a   tension~\cite{Ade:2015fva} between CMB and cluster estimation of the matter density $\Omega_M$
and the amplitude of density fluctuations $\sigma_8$. This may be due to an incorrect estimation of the total mass of the cluster, via the
mass-observable relation, in particular the scatter around the relation that may depend  on the redshift, the internal
structure and/or the dynamical state of the considered cluster. 
High-resolution SZ imaging of high-redshift cluster are thus needed to study their intra-cluster properties, such as dynamical states (mergers)
and morphology (departure from sphericity). It must be combined with other probes, in particular X-ray, within the
framework of multi-probe analysis of clusters of galaxies.

%

\section{SZ imaging of clusters of galaxies with the NIKA camera}

The NIKA camera has been used as a pathfinder for NIKA2, to demonstrate the possibility to use large arrays 
of KIDs in millimeter astronomy~\cite{Ritacco:2016due,Bracco}. In order to validate the use of a KID-based camera for SZ science, {\it i.e.} faint and diffuse signals, 
we have mapped the SZ signal in the direction of five clusters of galaxies 
~\cite{Adam:2013ufa,Adam:2014wxa,Adam:2015bba,Adam:2016abn,Ruppin:2016rnt,Adam:2017mlj} and 
combined with X-ray data to study the radial distribution of the thermodynamical properties of their intra-cluster medium, {\it i.e.}  pressure, density, temperature, 
and entropy radial profiles. This methodology is typically what is expected from the NIKA2 SZ large program and highlights  
the possibility of measuring thermodynamical profiles of high-redshift clusters, with a small integration time  
and without X-ray spectroscopy. The various SZ observations with NIKA, see tab. \ref{tab:nikaSZ}, are:

\noindent {\small $\bullet$} RX J1347.5-1145, a well-known
and strong SZ source,   allowed us 
to perform the first SZ cartography ever achieved with a KID-based camera~\cite{Adam:2013ufa}. In particular, we confirm that this cluster 
is an ongoing merger.

\noindent {\small $\bullet$} CL J1226.9+3332, a high-redshift cluster,     has been 
shown to present cluster parameters, mass and integrated Compton parameter,  consistent with the Planck best-fit 
scaling relation~\cite{Ade:2013lmv} 
obtained with a sample of nearby clusters. Although no conclusion can be drawn from a single high-redshift cluster, 
it highlights  the interest of the  
NIKA2 SZ large program that is dedicated to the observation of a cluster sample with redshift up to $0.9$.

\noindent {\small $\bullet$}   MACS J1423.9+2404 has been used to explore the impact of the presence of point sources~\cite{Adam:2015bba}.

\noindent {\small $\bullet$}  MACS J0717.5+3745 has been used to report 
the first mapping of the kinetic Sunyaev-Zel'dovich signal towards a cluster~\cite{Adam:2016abn} 
as well as the first mapping of the hot gas temperature using X-ray and SZ imaging~\cite{Adam:2017mlj}, 
providing an independant cross-calibration of X-ray spectroscopic measurements.

\noindent {\small $\bullet$} PSZ1 G045.85+57.71 has been used to demonstrate the possibility to use NIKA2 for a high-resolution follow-up 
of Planck-discovered clusters~\cite{Ruppin:2016rnt}. 
We have also proposed a new non-parametric deprojection procedure to extract the pressure profile, which has been shown to be in good
agreement with X-ray data, obtained with spectroscopy. The uncertainty on the integrated Compton parameter is reduced by a factor 2 with
respect to Planck result~\cite{Ruppin:2016rnt}. 

\begin{table}[h]
\begin{center}
\begin{tabular}{|c|c|c|c|}\hline 
               Cluster     & $z$ & Obs. time (h) & Ref. \\ \hline	 \hline      
	       RX J1347.5-1145 & $0.45$ & $5.5$ & \cite{Adam:2013ufa} \\
	       CL J1226.9+3332 & $0.89$ & $7.8$ & \cite{Adam:2014wxa,Romero:2017xri}\\
	       MACS J1423.9+2404 & $0.54$ & $1.5$ & \cite{Adam:2015bba} \\
	       MACS J0717.5+3745  & $0.55$ &$13.1$ & \cite{Adam:2016abn,Adam:2017mlj}  \\
	       PSZ1 G045.85+57.71 & $0.61$ & $4.3$  &\cite{Ruppin:2016rnt} \\ \hline
\end{tabular}
\caption{\it \small Overview of cluster observations with NIKA at the IRAM 30-m telescope.}
\label{tab:nikaSZ}
\end{center}
\end{table}

\section{The NIKA2 SZ large program}
The NIKA2 SZ large program is a follow-up of Planck- and ACT-discovered clusters and is part of the NIKA2 Guaranteed time. 
The NIKA2 cluster sample contains 50 clusters with redshift ranging between 0.4 and 0.9, selected from the 
Planck~\cite{Ade:2015gva,Ade:2013skr} and ACT catalogs~\cite{Hasselfield:2013wf}. This
representative sample will  be used for 
redshift evolution and cosmological studies. In particular, redshift bins present  an 
homogeneous coverage of cluster mass range as reconstructed from the integrated Compton parameter. 
The excellent  sensitivity of NIKA2 \cite{NIKA2-Adam} will allow us to obtain reliable SZ  
mapping of clusters of galaxies in only few hours (1 to 5 hours). 
The NIKA2 SZ data will be combined with ancillary data (X-ray, optical and radio). The study of the 
thermodynamic properties of the ICM  will lead to 
significant improvements on the use of clusters of galaxies 
to draw cosmological constraints. With the full sample, we aim at studying:

\noindent {\small $\bullet$} the thermodynamic properties of the cluster (temperature, entropy and mass radial profiles), 

\noindent {\small $\bullet$} the redshift evolution of the scaling law and of the cluster pressure profiles up to high redshift, 

\noindent {\small $\bullet$}  the cluster morphology and the dynamical state at high redshift 
(departure from spherical symmetry, merging events, cooling processes).
\section*{Acknowledgments}
{\small We would like to thank the IRAM staff for their support during the campaigns. 
The NIKA dilution cryostat has been designed and built at the Institut N\'eel. 
In particular, we acknowledge the crucial contribution of the Cryogenics Group, and 
in particular Gregory Garde, Henri Rodenas, Jean Paul Leggeri, Philippe Camus. 
This work has been partially funded by the Foundation Nanoscience Grenoble, the LabEx FOCUS ANR-11-LABX-0013 and 
the ANR under the contracts "MKIDS", "NIKA" and ANR-15-CE31-0017. 
This work has benefited from the support of the European Research Council Advanced Grant ORISTARS 
under the European Union's Seventh Framework Programme (Grant Agreement no. 291294).
We acknowledge fundings from the ENIGMASS French LabEx (R. A. and F. R.), 
the CNES post-doctoral fellowship program (R. A.),  the CNES doctoral fellowship program (A. R.) and 
the FOCUS French LabEx doctoral fellowship program (A. R.).}


\section*{References}
\begin{thebibliography}{99}

\bibitem{Calvo16}
M. Calvo {\it et al.},  
Journal of Low Temperature Physics 184 (2016) 816-823 

\bibitem{NIKA2-Adam}
 R.~Adam {\it et al.}, arXiv:1707.00908  

\bibitem{Monfardini11}    A. Monfardini {\it et al.}, The Astrophysical Journal Supplement 194 (2011)  24 
\bibitem{Bourrion12}    O. Bourrion {\it et al.}, JINST 11 (2016) P11001 
\bibitem{Calvo13}    M. Calvo {\it et al.}, Astron.\ Astrophys.\  {\bf 551} (2013) L12
\bibitem{Monfardini13} A. Monfardini {\it et al.}, JLTP 176 (2014) 787
    
 
\bibitem{Catalano:2014nml}
  A.~Catalano {\it et al.},
  Astron.\ Astrophys.\  {\bf 569} (2014) A9.
  



\bibitem{Sunyaev:1972}
R.~A.~Sunyaev and Y.~B.~Zel'dovich, Astrophys.~Space Phys.~Res.~ 4 (1972) 173
\bibitem{Sunyaev:1980vz}
  R.~A.~Sunyaev and Y.~B.~Zel'dovich,
  Ann.\ Rev.\ Astron.\ Astrophys.\  {\bf 18} (1980) 537.
  
\bibitem{Birkinshaw:1998qp}
  M.~Birkinshaw,
  Phys.\ Rept.\  {\bf 310} (1999) 97



 
\bibitem{Ade:2013lmv}
  P.~A.~R.~Ade {\it et al.} [Planck Collaboration],
  Astron.\ Astrophys.\  {\bf 571} (2014) A20



\bibitem{Ade:2015fva}
  P.~A.~R.~Ade {\it et al.} [Planck Collaboration],
  Astron.\ Astrophys.\  {\bf 594} (2016) A24


\bibitem{Ade:2015gva}
  P.~A.~R.~Ade {\it et al.} [Planck Collaboration],
  Astron.\ Astrophys.\  {\bf 594} (2016) A27
 
  
\bibitem{Ade:2013skr}
  P.~A.~R.~Ade {\it et al.} [Planck Collaboration],
  Astron.\ Astrophys.\  {\bf 571} (2014) A29

\bibitem{Hasselfield:2013wf}
  M.~Hasselfield {\it et al.},
  JCAP {\bf 1307} (2013) 008

 
\bibitem{Bleem:2014iim}
  L.~E.~Bleem {\it et al.} [SPT Collaboration],
  Astrophys.\ J.\ Suppl.\  {\bf 216} (2015) 2,  27







\bibitem{Ritacco:2016due}
  A.~Ritacco {\it et al.},
  Astron.\ Astrophys.\  {\bf 599} (2017) A34
  
\bibitem{Bracco}
  A.~Bracco {\it et al.},
  Astron.\ Astrophys.\  {\bf 604} (2017) A52


\bibitem{Adam:2013ufa}
  R.~Adam {\it et al.},
  Astron.\ Astrophys.\  {\bf 569} (2014) A66
  

\bibitem{Adam:2014wxa}
  R.~Adam {\it et al.},
  Astron.\ Astrophys.\  {\bf 576} (2015) A12
  


\bibitem{Adam:2015bba}
  R.~Adam {\it et al.},
  Astron.\ Astrophys.\  {\bf 586} (2016) A122
  
  
\bibitem{Adam:2016abn}
  R.~Adam {\it et al.},
  Astron.\ Astrophys.\  {\bf 598} (2017) A115



\bibitem{Ruppin:2016rnt}
  F.~Ruppin {\it et al.},
  Astron.\ Astrophys.\  {\bf 597} (2017) A110
  
\bibitem{Adam:2017mlj}
  R.~Adam {\it et al.},
  arXiv:1706.10230  
 
\bibitem{Romero:2017xri}
  C.~Romero {\it et al.},
  arXiv:1707.06113  
  
  
 








\end{thebibliography}
\end{document}